# β-phase manganese dioxide nanorods: Synthesis and characterization for supercapacitor applications

Veenu Kumari[1], Balram Tripathi[2] and Ambesh Dixit[1,*]

[1]*Indian Institute of Technology Jodhpur, Rajasthan, India 342011*

[2] *Dept. of Physics, S S Jain Subodh P G College Jaipur, India-302004*

*ambesh@iitj.ac.in

*Abstract*—Manganese dioxide nanorods were synthesized using novel solution route. The phase and microstructure of synthesized materials were identified using X-ray diffraction, scanning electron and transmission electron microscopic measurements. The material crystallizes into β crystallographic phase and consists of nanorods of diameter in the range of ~10 – 14 nm and length ~50 nm. The Fourier Transform Infrared (FTIR) spectroscopic and thermogravimetric analysis (TGA) measurements were carried out to understand the materials microscopic and thermal properties. The material exhibits characteristic Mn – O vibrational frequencies, confirming the phase purity of material. The electrochemical performance of β-$MnO_2$ nanorods was evaluated using cyclic voltammetry and galvanostatic charge/discharge measurements using in-house developed supercapacitor device assembly. We observed ~ 125 F/g specific capacitance for β- $MnO_2$ nanorods electrode materials.

*Index Terms*— β- $MnO_2$, nanorods, supercapacitor. Scanning electron microscopy, Tunneling electron microscopy

# I. INTRODUCTION

The energy storage is getting attention due to its potential for possible applications at the time of requirements. The research and development in the domain of energy storage has demonstrated in terms of high energy density rechargeable batteries and high power density supercapacitors. Lithium ion rechargeable batteries and carbon based supercapacitors are examples, commonly in use for everyday electronics. Yet the need of single device, capable of both high power density and high energy density simultaneously, is essential for power applications such as electric vehicles. The progress will rely on the development of novel electrode materials with properties such as higher operating voltage, storage capacity and power density; cater the needs [1 - 5]. There are continuous efforts for the development of such electrode materials exhibiting high operating voltage range with good storage capacity and fast charge/discharge rate. Supercapacitors with such electrode materials may show exceptionally long cycle life ≥ 100000 cycles in conjunction with enhanced storage capacity [5, 6]. In general, there are three types of electrode materials, carbon, conducting polymer and metal oxide based electrodes. The energy densities of supercapacitors based on these electrodes lie in the range of ~ 180 F/g, 220 F/g and 720 F/g respectively [2, 5]. Numerous rare earth and transition metal oxides such as $RuO_2$, $NiO$, $IrO_2$, $Co_3O_4$, $MnO_2$ etc have been explored for such applications [5 - 8]. The hydrous $RuO_2 \cdot xH_2O$ in amorphous geometry has received attention due to its very high specific capacitance ~720 F/g, yet the high materials cost, toxicity of ruthenium and low porosity restrict the material application for commercial applications [9 - 11]. To harness the high specific capacity of amorphous hydrous $RuO_2 \cdot xH_2O$, recent efforts are to develop the composites of $RuO_2 \cdot xH_2O$ with other less expensive metal oxides such as $MO_2$ (M = Mn, Pb, Ni etc) materials [12]. Several attempts are made to explore transition metal oxide materials such as hydrous $NiO$, $MnO_2$, $Co_3O_4$ etc for high specific capacity supercapacitors applications [13 - 15]. Manganese oxide among such oxide materials is important material because of its attractive high specific capacity, non-toxicity in conjunction with its earth abundance and environmentally compatible [7, 14]. This material has been explored with numerous electrolytes for its pseudocapacitance behavour where single electron transfer between $Mn^{4+}$ and $Mn^{3+}$ is responsible for the onset of high specific capacitance [5, 16]. Nevertheless, there are ambiguities about the specific capacitance of MnO2 in different crystalline phases and authors Deveraj et al [7] has reported that specific capacitance is largest in α and δ crystallographic phases ~ 250 F/g and lowest ~ 9 F/g for β crystallographic phase of $MnO_2$, where as some recent studies have shown that even nanostructured β $MnO_2$ material may exhibit better electrochemical performance [17]. We investigated the synthesis of single phase β $MnO_2$ nanorods and intensive material characterization and electrochemical measurements are carried out to understand the correlation of microstructure and electrochemical performance.

## II. EXPERIMENTAL

We carried out synthesis of β $MnO_2$ nanorods using novel solution route as suggested in Ref. 7 & 8. The synthesis process was modified and heat treatment was optimized for synthesis of single phase β $MnO_2$ nanords. The process consists of two starting solutions, called solution A and solution B, as explained in Fig. 1. The chemicals potassium permanganate ($KMnO_4$) and manganese (II) acetate ($Mn(CH_3COO)_2$) were procured from Alfa Aesar and used without any purification. Solution A was prepared by adding 0.1 M $KMnO_4$ in 40 ml of deionized (DI) water and solution B was prepared by adding 1.5 M $Mn(CH_3COO)_2$ in 4 ml of DI water separately. The homogeneous solutions were, then, mixed slowly under continuous stirring for 1 hour at 40 $^0$C. The brown precipitate was observed at completion of the chemical reaction, suggesting the formation of β $MnO_2$ nanorods. The precipitated solution was centrifuged about 7000 rpm to separate out the precipitated nanorods. These nanorods were collected and washed with DI water repeatedly till supernatant became neutral. The collected powder was dried at 80 $^0$C for 12 hours. The amorphous dried powder was heated at 400 $^0$C for 1 hr to convert the amorphous β nanorods into well oriented β crystallite phase. The diameter of the synthesized nanorods is in the range of ~10–14 nm and length ~50 nm.

The synthesized material was characterized using Powder X-ray diffraction (PXRD) from Bruker D8 Advanced X-ray diffractometer with Cu-K$_α$ (λ = 0.154 nm) as the incident radiation source for crystallographic information and phase identification. The PXRD data was recorded in θ/2 θ geometry for $20^0$ - $80^0$ range. The microstructure analysis was carried out using scanning electron microscopy (SEM) from Zeiss EVO 18 Sp. Ed and transmission electron microscopy (TEM) from JEOL-high voltage TEM systems. The energy-dispersive analysis by X-ray (EDAX) measurements were used to determine the elemental stoichiometric analysis of synthesized materials. The vibrational spectroscopic measurements are carried out using Bruker Ver.70v Fourier Transform Infrared spectrophotometer (FTIR). In addition, we also carried out thermogravimetric analysis (TGA) measurements using Perking Elmer TGA spectrometer under continuous flow of non reactive $N_2$ gas at the scanning rate of 10 $^0$C/min for temperature range of 30 $^0$C to 900 $^0$C.

We prepared 1"x1" working electrodes by mixing active β $MnO_2$ nanorods, carbon black and polyvinylidene fluoride (PVDF) in the weight ratio of 80%:20%:10% respectively and ground to form a homogeneous mixture. Here carbon black is used as a conducting material and PVDF as a binder in the β $MnO_2$ nanorod matrix. In addition, few drops of N-methyl-2-pyrrolidone (NMP) solvent were added in the mixture to form the slurry. The homogeneous slurry was, then, dispersed over the conducting Al foil and rolled over to form a uniform layer. The coated electrode material was dried in a conventional oven at 80 $^0$C for 10 hours in air. The dried Al foil coated with electrode material was cut into pieces of desired shape and size (1"x1") to form the supercapacitor assembly, as explained in Fig. 2. The average weight of ~ 40 mg was used as active electrode material for electrochemical analysis. The two electrode supercapacitor device was considered in this study. The device assembly consists of top and bottom covers, made of non reactive Perspex sheet, with embedded copper/aluminum thin sheets as current collector. The prepared electrodes are placed over copper current collectors, separated by a cellgaurd separator, as explained in the schematic device structure in Fig 2(a). The assembled copper/aluminum electrodes with Perspex sheet are shown in top panel in Fig 2(b) and the complete photograph of assembled device is shown in bottom photograph in Fig 2(b). To ensure the uniform electrical contact between current collector and electrode material, the complete assembly is tightened with two screws as shown in Fig. 2(b) bottom panel. The cyclic voltammetry (CV) and galvanostatic charge-discharge electrochemical measurements of these fabricated supercapacitor devices were carried out using electrochemical workstation; model CHI660E, at different scan rates. An aqueous electrolyte of 3M KOH was used for characterizing the assembled supercapacitor device and specific capacitance of the device was calculated using the following equation

$$C = I \cdot \Delta t / m \Delta V$$

Where C is the specific capacitance, I is the discharge current, $\Delta t$ is discharge time, m is the mass of active electrode materials, $\Delta V$ is the potential difference during discharge.

### III. Results and Discussions

The X-ray diffraction pattern of synthesized β MnO$_2$ nanorods is shown in Fig. 3. All peaks in this graph may be indexed to the crystalline β-MnO$_2$ structure (PDF # 812261) and consistent with the reported literature [7, 17]. The observed broad peaks in XRD pattern, shown in Fig 3(a), confirm the nanostructured geometry of β MnO$_2$ material, in consistent with our microscopic measurements, discussed later. The average grain size of β MnO$_2$ nanorods was calculated using Scherrer formula $d = k\lambda / \beta_{1/2} Cos\theta$, where k is dimesionless geometrical factor, λ is X-ray wavelength, $\beta_{1/2}$ is peak broadening in radians, and θ is the Bragg diffraction angle [18]. The (110), (101), (200) and (220) diffraction peaks from Fig. 3 are used for domain size calculation and observed size lie in the range of 10 – 14 nm, consistent with our electron microscopic measurements. The scanning electron micrographs are shown in Fig 4, explaining the large agglomerated grains, which are of random geometrical shapes. The ratio of elemental atomic fraction of manganese and oxygen, calculated from EDAX measurements, is ~ 32%:68%, explaining the stoichiometric composition of synthesized materials. Transmission electron micrographs of synthesized β MnO$_2$ nanorods are shown in Fig. 5. These micrographs exhibit the loosely packed MnO$_2$ nanorods with average diameter in the range of ~10 – 14 nm, consistent with the observed grain size from X-ray diffraction measurements. The average length of these nanorods is of the order of ~ 50 nm. The observed small diameter and length of these nanorods may play an important role to tailor the specific capacitance of the supercapacitor device. FTIR spectrograph is shown in Fig. 6 and the observed vibrational modes at ~ 713, 578 and 527 cm$^{-1}$ represent the characteristic Mn-O vibrations modes in Mn-O octahedral coordination and consistent with reported literature [13]. The weightloss versus temperature measurement for β $MnO_2$ nanorods was carried out using TGA measurements and results are plotted in Fig. 7. The weightloss versus temperature graph clearly shows two different regions at ~ 100 $^0C$ and 580 $^0C$. The slope near ~ 100 $^0C$ may represent the burn outs of organics and physically absorbed water, left in the material during synthesis, (not observed in bulk $MnO_2$ materials, as shown in inset), whereas slope change at ~ 580 $^0C$ is consistent with bulk $MnO_2$ measurement, as shown in inset, may represent the dissociation of oxygen from $MnO_2$ representing an endothermic process. The high temperature dissociation of $MnO_2$ suggests that this material may exhibit stable electrochemical performance over a wide range of temperature, without degrading the device performance.

The electrochemical performance of in house designed supercapacitor was determined using cyclic voltammetry and galvanostatic charge/discharge measurements at room temperature. These measurements were carried out within the potential window of 0 – 0.8 V at different scan rates of 100, 300 and 500 mV/s. 3M KOH electrolyte solution was used for all these measurements. The observed cyclic voltammograms for β $MnO_2$ nanorods at different scan rates are shown in Fig. 8. The rectangular shape of these voltammograms represent the characteristic feature of electric double-layer capacitor [2] and increasing scan rate suggest higher electric double-layer capacitance with enhanced cyclic reversibility and stability for charge/discharge process. All the measurements were carried out for 8 cycles to understand the fading of stored charge at different scan rates, we do not observe any degradation under investigation, as evident from these voltammograms. The absence of any peak in these voltammograms (Fig. 8) suggests the intrinsic nature of electrostatic mechanism of charge storage for supercapacitor and charge/discharge is purely electrostatic and free from any redox reaction. The similar electrostatic behavior was observed with higher scan rates, suggesting that β $MnO_2$ based supercapacitor may exhibit stable performance at elevated charge/discharge rates. The calculated specific capacitance [$C_{sp} = 2\ i/S*m$), where $i$ is the current, $S$ scan rate and $m$ mass of electrode] at different scan rates is plotted in Fig 8. The galvanostatic charge/discharge measurements are carried out in the potential window -0.8 – 0.8 V at 0.1 A constant current and graphs are shown in Fig. 9.

The measurements are quasi triangular in nature, representing the electrostatic nature of charge discharge nature for fabricated supercapacitor device. The large charge/discharge time, as evident in Fig. 9, suggests that large number of electrons and electrolyte ions are participating in the charge/discharge process of β $MnO_2$ nanorod based electrodes. A sharp drop has been observed in the discharge curves (Fig. 9). The onset of this drop in the discharge is the consequence of diffusion-limited mobility of the electrolyte ions in the electrode pores and such limitations are associated with the equivalent series resistance (ESR) of the supercapacitor device and we found ~160 mΩ ESR values for fabricated supercapacitor devices. The calculated specific capacitance [$C_{sp} = 2\ i/((dV/dt)*m)$ where $i$ is the current used during charge/discharge, $dV/dt$ the potential gradient and $m$ the mass of electrode material] using GCD data, is ~ 125 F/g for as prepared β $MnO_2$ nanorod electrode based supercapacitors, consistent with specific capacitance calculated from voltammograms (Fig 8).

## IV.  CONCLUSIONS

In conclusion, we successfully synthesized β $MnO_2$ nanorods and investigated their physical properties in conjunction with electrochemical performance. High voltage supercapacitor devices were successfully fabricated using synthesized β $MnO_2$ nanorods electrode material. We observed that β $MnO_2$ nanorods based supercapacitor devices exhibit better electrochemical performance and calculated specific capacitance is ~125 F/g at 1A/g of charge/discharge current rate, much higher than that of reported values [7], but lower than theoretical specific capacitane ~1100 F/g [5]. These studies provide possibility for use of β $MnO_2$ as a high voltage and high energy storage supercapacitor devices, which may work at high charge/discharge rate with enhanced specific capacity. In addition, the materials thermal stability in nanostructured geometry, suggest that β $MnO_2$ nanorods based supercapacitor devices may work at much higher temperature without hampering the device performance.

## IV. ACKNOWLEDGMENT

This work was supported by the MNRE through grant ECESTRE/20110007. We acknowledge helpful discussion with Laltu Chandra and Tapano Hotta.

Figure 1: Schematic process flow for synthesis of β MnO$_2$ nanorods, using solution route

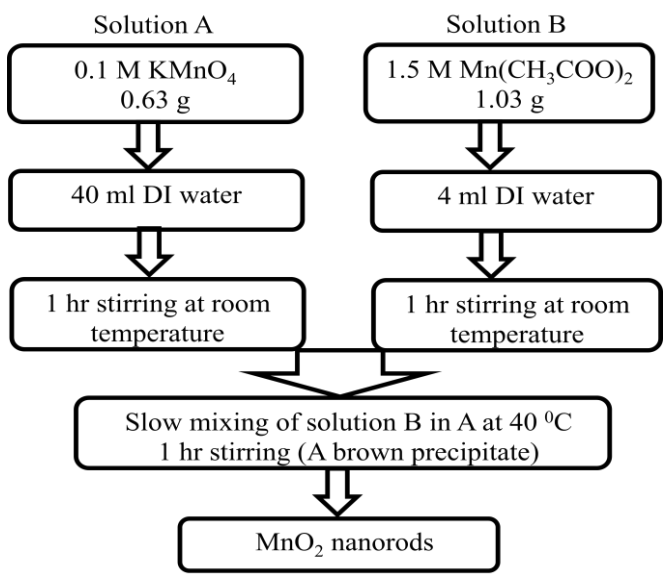

Figure 2(a) Schematic representation of different components of supercapacitor assembly; (b) actual photographs of top and bottom electrodes with current collectors (top panel) and assembled device (bottom panel) (Color online)

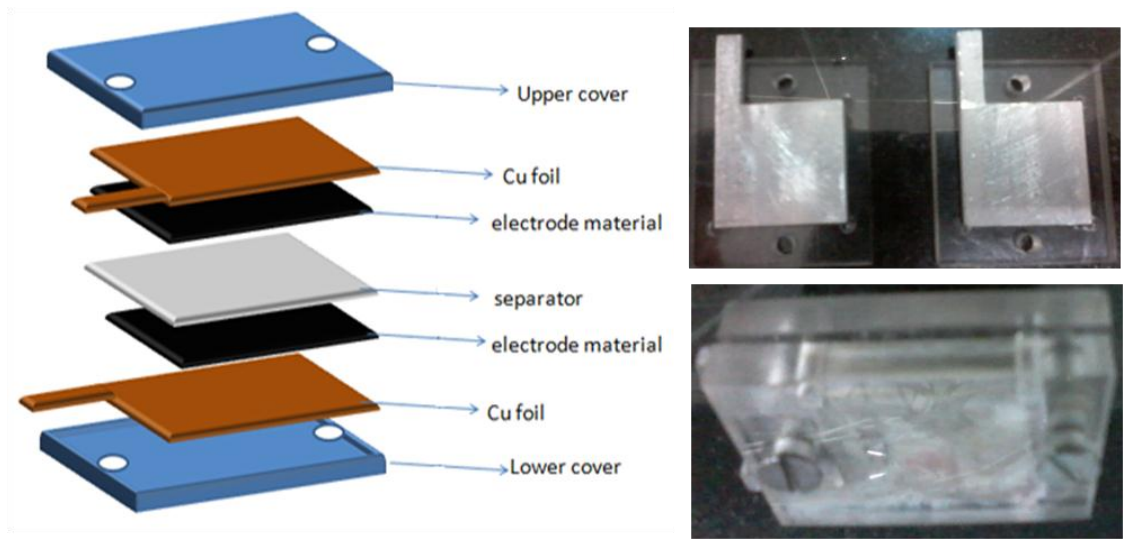

Figure.3 Powder X-ray diffraction graph of β-MnO₂ nanorods, after air annealing at 400 °C

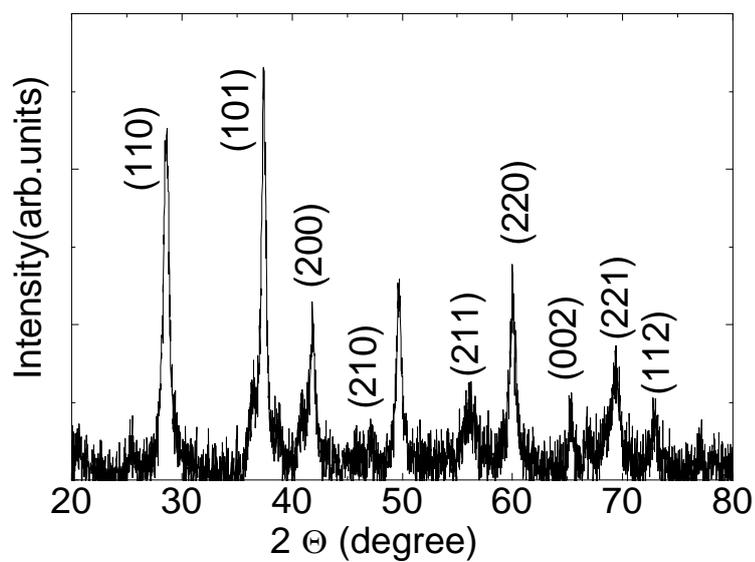

Figure 4 SEM micrographs for β MnO₂ nanorods powder sample, with inset showing EDX spectrum for elemental Manganese and Oxygen analysis (Color online)

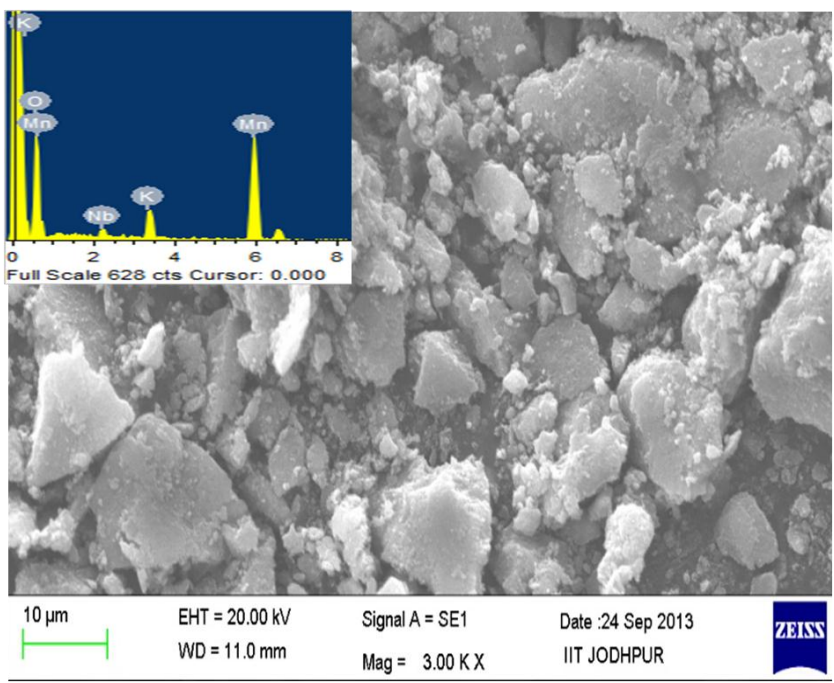

Figure 5 Transmission electron micrographs of as prepared β MnO$_2$ nanorods at 50 nm (a) and 100 nm (b) length scales

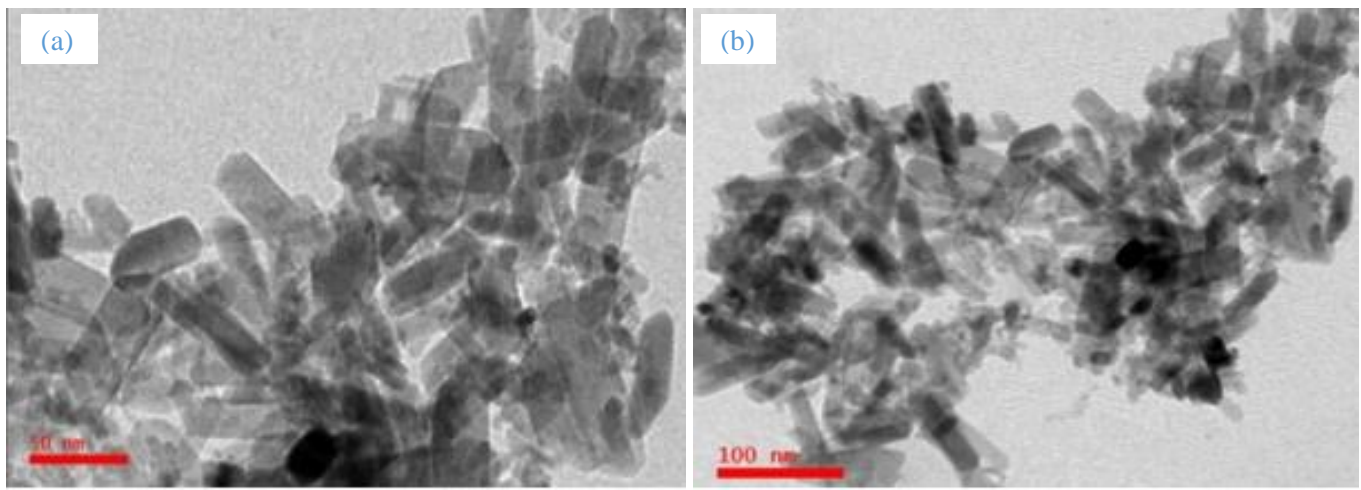

Figure 6 Fourier transform infrared spectrum (FTIR) of as prepared β MnO$_2$ nanorods, showing characteristic features of Mn-O vibrations. Feature at 1500 cm$^{-1}$ is noise due to ambient nitrogen and other residual gases present during the experiment.

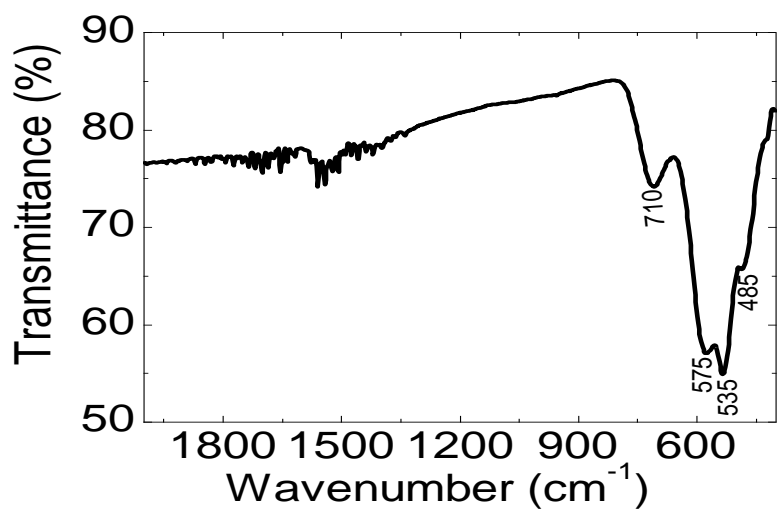

Figure 7 TGA graph of as prepared β MnO$_2$ nanorods, with inset showing TGA graph for bulk β MnO$_2$ powder sample.

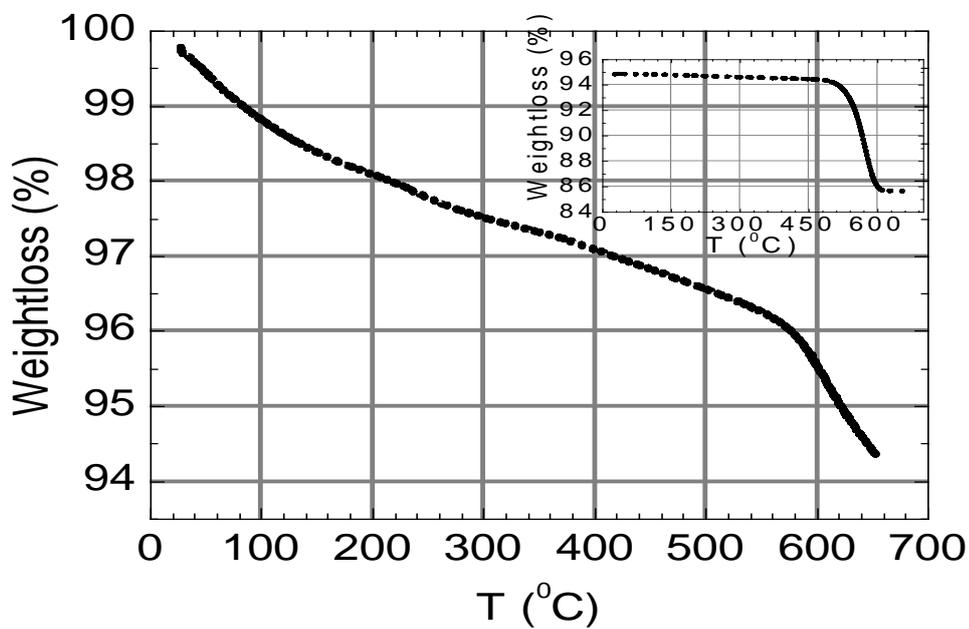

Figure 8 Current –Votage voltammograms for β MnO$_2$ nanorods as active electrode material in 2 electrode supercapcitor assembly at different scan rates (Color online)

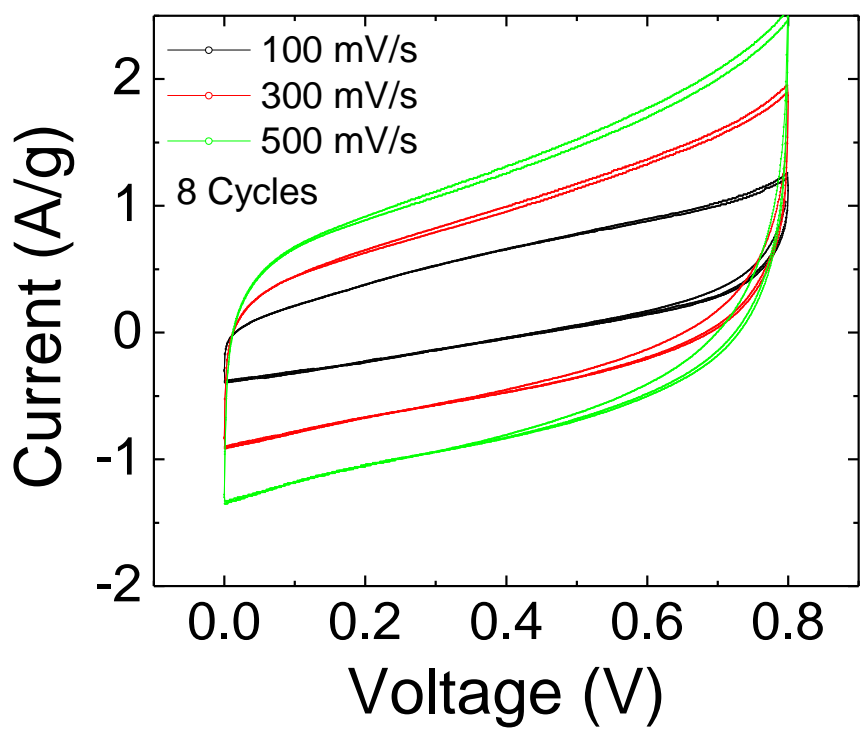

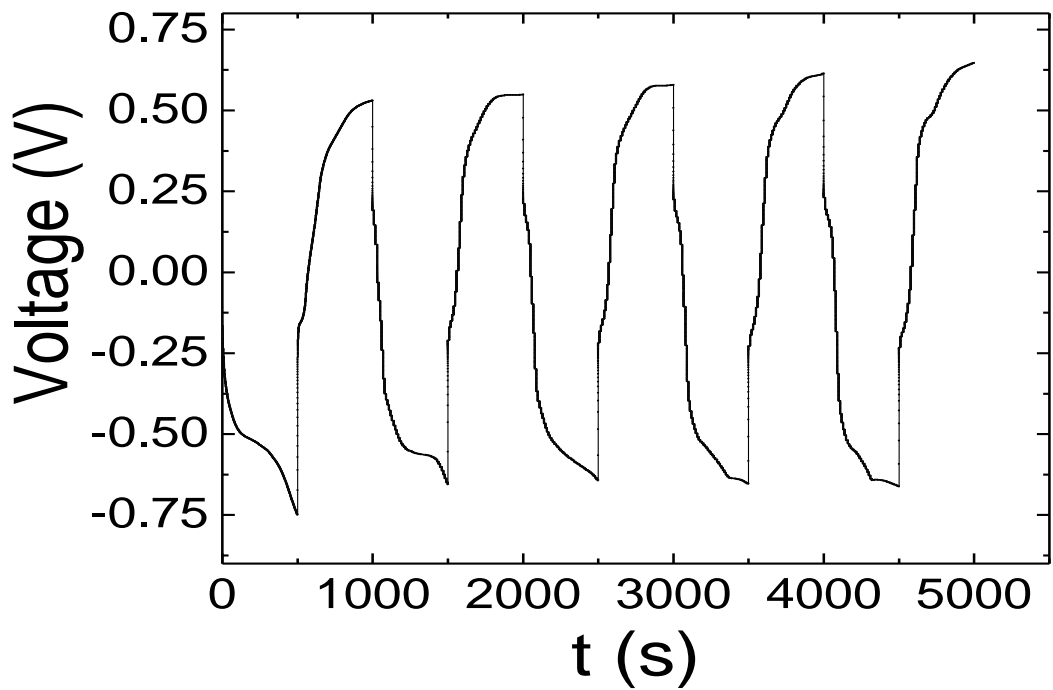

Figure 9 Galvanostatic charge/discharge graphs for β MnO$_2$ nanorod electrodes at 0.1 A constant current.